\documentclass[%
 reprint,
amsmath,amssymb,
aps,
prl,
floatfix,
]{revtex4-1}

\usepackage{graphicx}
\usepackage{dcolumn}
\usepackage{bm}
\usepackage{color}

\begin{document}

\preprint{APS/123-QED}

\title{The Reduction of Magnetic Reconnection Outflow Jets to Sub-Alfv\'enic Speeds}

\author{Colby C. Haggerty}
\email{CHaggerty@UChicago.edu}
\affiliation{Department of Astronomy and Astrophysics, 
            University of Chicago, Chicago, IL 60637, USA}
\affiliation{Bartol Research Institute, Department of Physics and
			 Astronomy, University of Delaware, Newark, DE 19716, USA }
            
\author{Michael A. Shay}%
\author{Alexandros Chasapis}%
\affiliation{Bartol Research Institute, Department of Physics and
			 Astronomy, University of Delaware, Newark, DE 19716, USA }
             
\author{Tai D. Phan}
\affiliation{Space Sciences Laboratory, University of California, Berkeley, California 94720, USA}

\author{James F. Drake}
\affiliation{Institute for Research in Electronics and Applied Physics, University of Maryland, College Park, Maryland 20742, USA}

\author{Kittipat Malakit}
\affiliation{Department of Physics, Faculty of Science and Technology, Thammasat University, Pathum Thani, Thailand}

\author{Paul A. Cassak}
\affiliation{Department of Physics and Astronomy, West Virginia University, Morgantown, West Virginia 26506, USA}

\author{Rungployphan Kieokaew}
\affiliation{CGAFD, Mathematics, CEMPS, University of Exeter, Exeter, UK}

\date{\today}

\begin{abstract}
The outflow velocity of jets produced by collisionless magnetic reconnection is shown to be reduced by the ion exhaust temperature in simulations and observations.
We derive a scaling relationship for the outflow velocity based on the upstream Alfv\'en speed and the parallel ion exhaust temperature, which is verified in kinetic simulations and observations. The outflow speed reduction is shown to be due to the firehose instability criterion, and so for large enough guide fields this effect is suppressed and the outflow speed reaches the upstream Alfv\'en speed based on the reconnecting component of the magnetic field.
\end{abstract}

\maketitle


\section{\label{sec:Intro} Introduction}
Magnetic reconnection is a plasma process that efficiently releases 
energy stored in magnetic fields and generates fast plasma jets 
\citep{Priest2000MagneticApplications}. Reconnection occurs at 
thin current sheets where the direction of the magnetic field 
changes over a small spatial scale. During reconnection, 
field lines effectively break and cross connect, creating 
stretched field lines, which relax via the tension force. This contraction of the field line generates bulk plasma outflow away 
from the location where this breaking occurred (called the x-line) with a speed expected to reach the Alfv\'en 
speed based on the inflowing plasma parameters and the changing 
component of the 
magnetic field \citep{Parker1957SweetsFluids}. 

Outflow jets 
are one of the most recognizable features of reconnection. They have been repeatably 
observed in simulations as well as laboratory and satellite observations 
(e.g., \citet{Paschmann1979PlasmaReconnection, 
Sonnerup1981EvidenceMagnetopause, Sato1979ExternallyConverter, 
Birn1981Three-dimensionalTail, Stenzel1982MagneticFlow}). Observational events typically require a jet detection to be classified as reconnection. 
However, in both simulations and observations, the magnitude of the outflow 
speed is often found to be significantly less than the Alfv\'en speed 
(e.g., \citet{Paschmann1986TheObservations,Phan1996Low-latitudeReconnection, Gosling2007DirectSheet, Liu2012TheBoundary, Haggerty2015TheReconnection}). A significant amount of the converted magnetic energy is transfered into the outflow jets \citep{Drake2009IonExhausts,Eastwood2013EnergyMagnetotail,Lapenta2014ElectromagneticReconnection,Yamada2014ConversionPlasma,Shay2014ElectronStudy,Phan2013ElectronShear} and the magnitude of theses jet velocities have important consequences for many different collisionless plasma systems where reconnection occurs (e.g. dipolarization fronts \citep{Fu2013DipolarizationEvidence, Pritchett2015ReconnectionFronts}, in thermal \citep{Drake2006FormationReconnection, Haggerty2015TheReconnection} and non-thermal particle energization \citet{Dahlin2014TheReconnection,Dahlin2015ElectronField} as well as potential many others). The ability to accurately predict the outflow velocity or reconnection jets is a fundamental and critical step in a complete description of magnetic reconnection.


In this work, we show that the firehose instability criterion being reached in the exhaust \citep{Liu2011TheTheory,Liu2012TheBoundary} reduces the 
outflow velocity in nearly anti-parallel reconnection events and derive a 
prediction for the reduction . A relationship is derived by matching the 
anisotropic Rankine-Hugoniot conditions across the edge of the reconnection exhaust boundary. The prediction is tested using 81
particle-in-cell (PIC) reconnection simulations and 14 previously
published observational events and is found to agree remarkably well. 
Finally, we discuss the implications of this result and its importance to 
understanding how released magnetic energy is partitioned during reconnection.

\section{\label{sec:Sim} Simulations and Observations}
To study why the outflow velocity is less than the Alfv\'en speed, we examine 81 reconnection simulations (56 where the two reconnecting field lines are separated by more than $135^{\circ}$ (nearly anti-parallel) and 25 where they are separated by some smaller angle (in this work $< 135^{\circ}$) (guide field)) preformed using the kinetic-PIC code P3D \citet{Zeiler2002Three-dimensionalReconnection}. In the simulations, magnetic field strengths and particle number densities are normalized to arbitrary characteristic values \(B_{0}\) and \(n_{0}\), respectively.
 Lengths are normalized to the ion inertial length \(d_{i0}=c/\omega_{pi0}\) at the reference density \(n_{0}\). Time is normalized to the ion cyclotron time \(\Omega_{ci0}^{-1}=(eB_{0}/m_{i}c)^{-1}.  \) Speeds are normalized to the Alfv\'en speed \(c_{A0}=\sqrt{B_{0}^{2}/(4\pi\,m_{i}\,n_{0})}\). Electric fields and temperatures are normalized to \(E_{0}=c_{A0}B_{0}/c\)  and \(T_{0}=m_{i}c_{A0}^{2}\), respectively. The coordinate system is in ``simulation coordinates,'' meaning that the reconnection outflows are along \(\hat{x}\) and the inflows are along \(\hat{y}\). 

Simulations are performed in a periodic domain with size and grid scale varied based on simulation and inflow parameters. The reconnection simulation parameters are described in detail in two previous publications \cite{Shay2014ElectronStudy,Haggerty2015TheReconnection}
A range of reconnection magnetic fields $B_r$, upstream densities $n_{up}$, and upstream
ion and electron temperatures $T_{i,up}$ and $T_{e,up}$ are used. The parameters for the simulations are shown in the Supplementary Material. 

In each simulation we take a trapezoidal region from $5$ to $20\, d_i$ downstream of the x-line bounded by the exhaust boundary in order to calculate the average parallel ion exhaust temperature. The outflow velocity is taken as the asymptotic $E\times B$ velocity at the midplane sufficiently far downstream of the x-line. Further details about the calculation of these values are detailed in \citet{Haggerty2015TheReconnection}. 
\begin{figure}
\begin{center}
\includegraphics[width=3.4in]{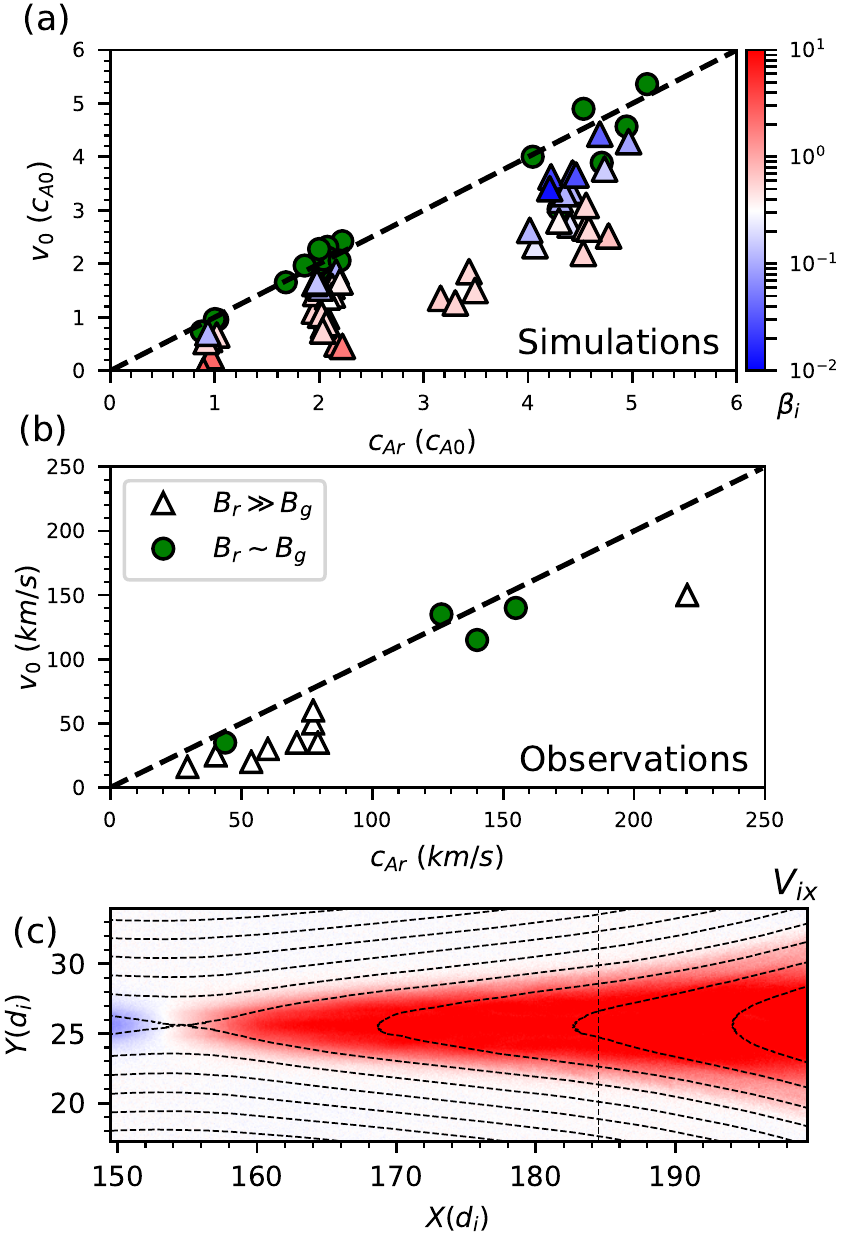}
\caption{(a \& b) The asymptotic $E\times B$ outflow velocity $v_0$ against the upstream Alfv\'en speed based on the reconnecting magnetic field $c_{Ar}$ for 81 different PIC simulations (a) and 14 previously published observational events (b). The green circles correspond to simulations with a guide field comparable to the reconnecting field, and the blue/white/red triangles correspond to nearly anti-parallel reconnection events. (c) 2D color plot of the ion outflow velocity $v_{ix}$ for an example simulation with contours of the magnetic field plotted in black dashed lines. The vertical dashed black line shows where the cut is taken for Fig.~\ref{Fig:2}.  This is simulation 693 in the table in the Supplementary Material.}
\label{Fig:1}
\end{center}
\end{figure}


Along with simulations we also examine 14 previously published observed reconnection events \citep{Gosling2005MagneticSheet, J.T.Gosling2006Petschek-typeUlysses, Gosling06, Davis2006,  Phan2007EvidenceMagnetosheath, Gosling2007DirectSheet, Phan2009PrevalenceAU, Hietala2015IonJet, ieroset2016MMSMagnetopause, ieroset2017THEMISfield}. These events were measured in several different plasma systems, including the solar wind, the magnetosheath and the magnetotail.

In Fig.~\ref{Fig:1}a and b the asymptotic $E\times B$ outflow velocity is plotted against the upstream Alfv\'en speed based on the reconnecting magnetic field $c_{Ar}$ with a dashed black line corresponding to a slope of 1 for the simulations and observations respectively. For the guide field cases (green circles) there is good agreement between the outflow velocity and the upstream Alfv\'en speed based on the reconnecting magnetic field. However, for every nearly anti-parallel simulation and observation (red white and blue triangles) irrespective of the initial conditions, the outflow velocity is less than the Alfv\'en speed. The difference between the outflow and the Alfv\'en speed also varies dramatically between different events. There are numerous cases where the outflow is almost as large as  Alfv\'en speed, and there are also many events where the outflow is an order of magnitude smaller than the Alfv\'en speed. It is clear that in the absence of a guide field, the outflow velocity can be significantly reduced. 

The reduction of the outflow velocity is linked in some way to the upstream ion beta $\beta_i$ based on the reconnecting component of the upstream magnetic field indicated by the triangle's color in Fig.~\ref{Fig:1}a. As $\beta_i$ goes to zero (blue triangles), the outflow approaches the Alfv\'en speed and as $\beta_i$ becomes larger (red triangles) the outflow velocity is reduced. This suggests that the ion upstream thermal velocity relative to the upstream Alfv\'en speed reduces the outflow. This is explored in the following section where we analyze the effects of the firehose instability in reconnection exhausts. In the next section, a theory for this relationship is derived.

\section{\label{sec:Body} Theory}
Using double adiabatic theory \citep{G.F.Chew1956BoltzmannCollisions} for a parallel propagating shear Alfv\'en wave , the dispersion relationship can be shown to be $\omega^2 = \frac{k^2}{m_in}( \frac{B^2}{4\pi} + P_\perp - P_\parallel)$ where $P_\parallel$ and $P_\perp$ are the sum of the ion and electron thermal pressures parallel and perpendicular to the local magnetic field line\citep{Parker1958DynamicalDensity}. For an isotropic system the relationship becomes $\omega = c_A k$ with the phase velocity equal to the Alfv\'en speed. There is clearly a regime where the right side is negative and the wave is unstable. This occurs when $P_\parallel > \frac{B^2}{4\pi} + P_\perp$ or equivalently when $\epsilon$ defined as $\epsilon = 1 + 4\pi (P_\perp - P_\parallel)/B^2 < 0$ and is referred to as the firehose instability. Conceptually, this instability can be interpreted as an effective centrifugal force from particles traveling along curved magnetic field lines due to $P_\parallel$ that beats the tension force trying to straighten out the field line. When $\epsilon \rightarrow 0$, there is no tension in the magnetic field line and the field line can not accelerate the plasma.

\begin{figure}
\begin{center}
\includegraphics[width=3.2in]{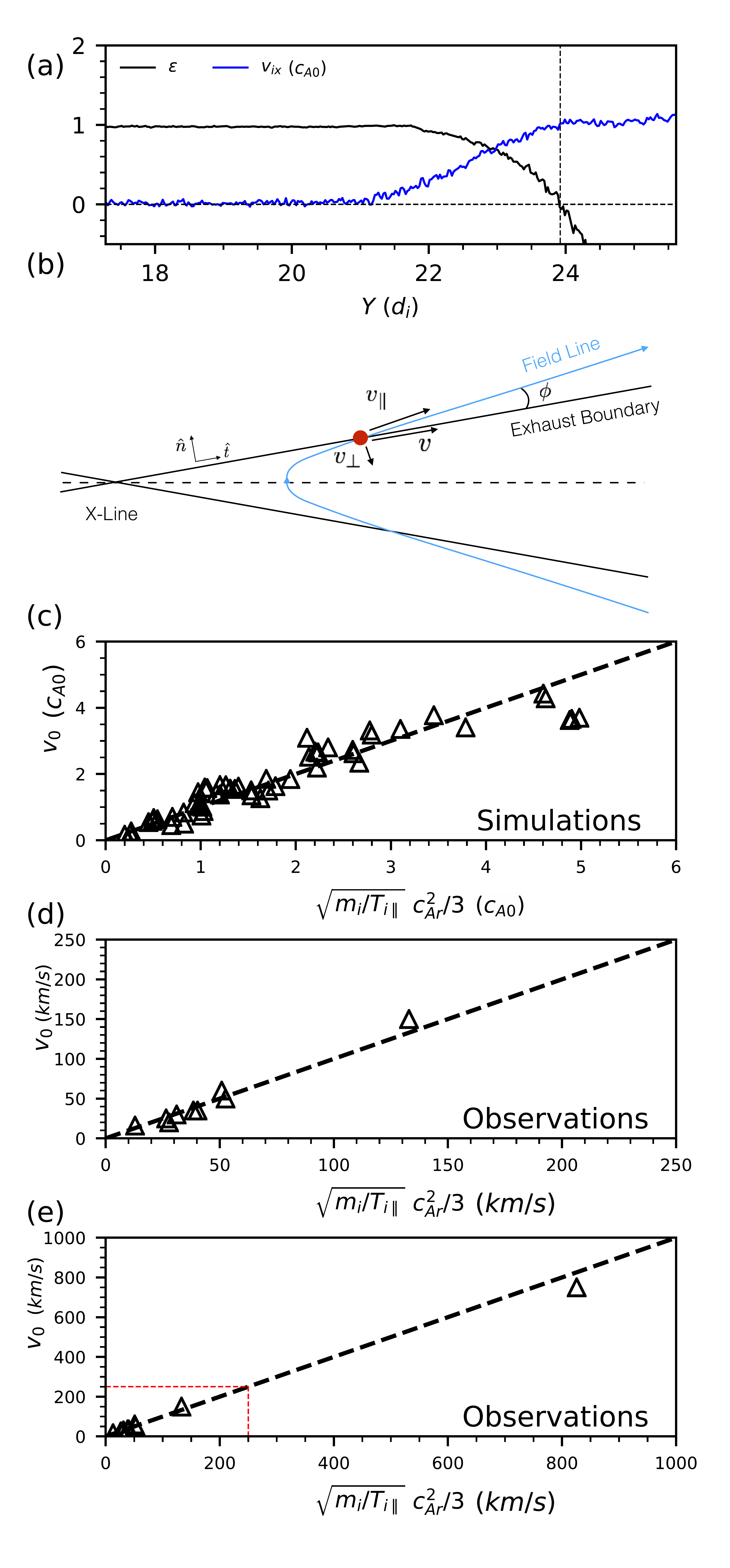}
\caption{(a) A cut of the ion outflow velocity $v_{ix}$ and the firehose parameter $\epsilon$ along the y direction in simulation 693. This cut is taken $30 d_i$ downstream of the x-line and is shown in Fig.~\ref{Fig:1}c. (b) Diagram of an accelerated ion's (red circle) motion relative to the exhaust boundary (black line) and the magnetic field (blue curve). (c) - (e) The asymptotic $E\times B$ outflow velocity versus the outflow prediction described in Eq.~\ref{Eq:prediction} for PIC simulations (c) and for observations (d) and (e) for nearly anti-parallel events. Panel (e) is the same as (d) but it includes a magnetotail event with an outflow velocity so large that it obscures the data from the other events. The red dashed box shows the limits of panel (d) for comparison.}
\label{Fig:2}
\end{center}
\end{figure}

The effect of the firehose instability in the exhaust can be seen in Fig.~\ref{Fig:2}a which shows a cut of the ion outflow velocity and the firehose parameter along the inflow direction at the location shown in Fig.~\ref{Fig:1}c. Precisely at the region where $\epsilon \rightarrow 0$, the outflow velocity stops increasing and flattens off. This suggests that the firehose instability is potentially responsible for limiting the outflow velocity.

To derive a prediction for the scaling of the outflow velocity we analyze the anisotropic Rankine-Hugoniot jump conditions across the reconnection exhaust boundary layer \citep{HudsonP1970DiscontinuitiesWind}. Note that the shock boundary is different from the separatrix which is the topological boundary defined by the field line passing through the x-point; the exhaust boundary is a line quasi-parallel to, but contained within, the separatrix which separates the hot, fast flowing exhaust plasma and the cooler, slowly inward convecting plasma (diagrammed in Fig.~\ref{Fig:2}b). The jump conditions are: \citep{Liu11b,Liu12,HudsonP1970DiscontinuitiesWind}:
\begin{align}
[B_n]_d^u = 0\\
[nv_n]_d^u = 0\\
\left[m_i n v_n \bm{v}_t - \epsilon\frac{B_n\bm{B}_t}{4\pi} \right ]_d^u = 0,
\end{align}
where $\left [... \right ]_d^u$ represent difference between the upstream and downstream of the shock and the subscripts $n$ and $t$ correspond to the directions normal and tangential to the shock, respectively, shown in Fig.~\ref{Fig:2}b. 
By comparing at a point upstream of the shock region and the location where $\epsilon$ goes to zero in the exhaust, we find from Eq.~1 and 3 that the outflow velocity (which at this location $v_0 \approx v_{td}$) satisfies the relationship
\begin{equation}\label{Eq:RH}
(n v_{n})_d v_0 = (n v_{n} v_t)_u - \epsilon_u\frac{B_nB_{tu}}{4\pi m_i}.
\end{equation}
This equation can be further simplified by noting that the upstream inflowing velocity $v_{nu}$ and the upstream flow tangential to the exhaust boundary $v_{tu}$ are both very small quantities compared to either the outflow velocity or the upstream Alfv\'en speed. The first term of the right hand side of Eq.~\ref{Eq:RH} can therefore be ignored and  the outflow velocity becomes
\begin{equation}\label{Eq:RHv0}
v_0 = - \epsilon_u\frac{B_nB_{tu}}{4\pi m_i n_u v_{nu}},
\end{equation}
where we have used $n_dv_{nd} = n_uv_{nu}$ from Eq.~2.

Fig.~\ref{Fig:2}b shows a schematic of a magnetic field line threading through the exhaust boundary layer at a small angle $\phi$. The magnetic field and bulk flow can then be rewritten in terms of this angle:
\begin{align}
B_{tu} &= B\cos{\phi} \approx B_r,
\label{Eq:B_u1}\\
B_n &= B\sin{\phi} \approx B_r \phi, \\
v_{nu} &= -v_{E\times B}\cos{\phi} \approx -v_{E\times B},
\label{Eq:v_u1}
\end{align}
where $v_{E\times B}$ is the $E\times B$ velocity of the inflowing plasma. Substituting Eqs.~\ref{Eq:B_u1}~-~\ref{Eq:v_u1} into Eq.~\ref{Eq:RHv0}, we find:
\begin{equation}\label{Eq:RHphi}
v_0 = \epsilon_u c_{Ar}^2\frac{\phi}{v_{E\times B}}
\end{equation}

In Fig.~\ref{Fig:2}b we show a diagram of an ion population (denoted by the red circle) that has been accelerated by the Fermi mechanism and is now traveling out of the exhaust along a field line. In the magnetic field coordinate system, the ion's velocity a parallel $v_\parallel$ and perpendicular $v_\perp$ component. This population of ions entered the the reconnection exhaust closer to the x-line and is now traveling away from the midplane on the upper half of the exhaust. These ions mix with the inflowing plasma and form the density enhancement in the exhaust associated with the exhaust shock boundary. Since these ions make up the shock boundary, the populations velocity should be parallel to the shock. This implies that $\phi \approx \tan{\phi} = v_\perp/v_\parallel$.
Using the $E \times B$ velocity for the perpendicular velocity and substituting this into Eq.~\ref{Eq:RHphi} we find the outflow velocity should be $v_0 = \epsilon_u c_{Ar}^2/v_\parallel$
where $v_\parallel$ is the parallel velocity component of the ions flowing away from the midplane at the leading edge of the exhaust boundary. 

It is not clear exactly what value $v_\parallel$ should have. In the limit where the inflowing ion thermal velocity is much larger than the Alfv\'en speed, only half of the inflowing population would enter the exhaust (the half with a parallel velocity pointing towards the midplane) and then $v_\parallel \sim \sqrt{T_i/m_i}$. In this limit as the upstream ion temperature increases, the outflow velocity becomes much smaller than the Alfv\'en speed. In the limit of cold inflowing ions, if we neglect the effect of the potential associated with electron trapping \citep{Le2009EquationsReconnection,Egedal2008EvidenceReconnection,Haggerty2015TheReconnection}, the parallel velocity is simply be $v_\parallel = 2v_0$. Using this, the outflow velocity should be $v_0 = c_A/\sqrt{2}$. This can be interpreted as an upper bound for the outflow velocity in anti-parallel reconnection. Using this prediction for the outflow with the ion heating predicted by Eq.~8 in \citet{Drake2009IonExhausts}, the total ion heating is $\Delta T_i = .167 m_i c_A^2$ which is within 20\% of the reported heating identified in observations and  simulations \citep{Haggerty2015TheReconnection,Phan2014IonAngle}. The behavior in both of these limits and the physical interpretation of this velocity suggests that $v_\parallel \propto \sqrt{T_{i\parallel}/m_i}$, where $T_{i\parallel}$ is the parallel ion temperature in the exhaust. Note that this temperature is the parallel ion exhaust temperature and thus includes both the initial upstream temperature as well as the additional temperature generated during reconnection. Furthermore, in the cold upstream, potential free limit $T_{i\parallel} \sim \Delta T_{i\parallel} = v_\parallel$, and in the hot upstream limit $T_{i\parallel} \sim T_{i,up} \sim v_\parallel$. From this we find 
\begin{equation}\label{Eq:Vprop}
v_0 \propto \epsilon_u \frac{c_{Ar}^2}{\sqrt{T_{i\parallel}/m_i}}
\end{equation}

Because $T_{i\parallel}$ includes the ion heating generated during reconnection and the heating is linked to the outflow velocity \citep{Drake2009IonExhausts,Haggerty2015TheReconnection} this relationship does not uniquely determine the outflow velocity. 
Eq.~\ref{Eq:Vprop}, however, provides an important link between temperature and exhaust velocity which can be tested experimentally and numerically. It may ultimately form the basis of a complete predictive theory for both outflow velocities and heating.
This is an important outstanding problem in reconnection and plasma physics and should be addressed in the future.

To test Eq.~\ref{Eq:Vprop} we examine nearly anti-parallel simulations and observations shown in Fig.~\ref{Fig:1}a-b. In Fig.~\ref{Fig:2} the outflow velocity measured in nearly anti-parallel simulations (c) and observations (d and e) is plotted against the formula given in Eq.~\ref{Eq:Vprop} using $1/3$ as the proportionality constant. The agreement between the theory and measured outflow velocity is remarkable. Both simulation and observation data points now lie along a single line with the same empirical factor of $1/3$ for both, whereas before the inclusion of the ion temperature term events would have a range of different velocity for the same prediction (i.e. the spread in $y$ in Fig.~\ref{Fig:1} a).

In Fig.~\ref{Fig:2}e the same data are shown as in Fig.~\ref{Fig:2}d with an extra observational event. This event occurred in Earth's magnetotail and was studied extensively by \citet{Hietala2015IonJet}. The magnitude of the upstream Alfv\'en speed in this event is so large that is dwarfs all the other events. The red dashed lines show the limits of Fig.~\ref{Fig:2}d to emphasize this point. In this event, the outflow velocity was significantly reduced from the Alfv\'en speed by as much as $400 km/s$; its speed is consistent with the theory presented here. 
Putting in the empirical multiplicative factor of 1/3 gives an accurate prediction for the outflow velocity $v_0$ in nearly anti-parallel symmetric magnetic reconnection:
\begin{equation}
v_0 = \frac{\epsilon_u}{3}\frac{c_{Ar}^2}{\sqrt{T_{i\parallel}/m_i}},
\label{Eq:prediction}
\end{equation}
and in the presence of a sufficiency strong guide field, the outflow speed is
\begin{equation}
v_0 = c_{Ar}.
\label{Eq:gf_prediction}
\end{equation}
The outflow velocity reaching the Alfv\'en speed in guide field reconnection is consistent with the physics leading to Eq.~\ref{Eq:Vprop}. In the presence of a strong guide field, the firehose instability is suppressed in the reconnection exhaust, leaving the reconnected field lines' tension force intact.

Lastly, we estimate the strength of the guide field required to transition from firehose unstable to stable. The firshose instability will be suppressed when the guide field in the exhuast is large enough to keep $\epsilon \geq 0$. Neglecting the reconnecting component of the magnetic field in the exhaust and compressional effects in the exhaust, this translates to $\frac{B_g^2}{8\pi n} \geq \Delta T_{i\parallel} - \Delta T_{i\perp}$. The difference in heating can be estimated from observations and simulations where the total ion heating is found to be $(\Delta T_{i\parallel} + 2\Delta T_{i\perp})/3 \sim 0.125 \frac{B_{r}^2}{4\pi n}$ and $\Delta T_{i\parallel} \approx 2\Delta T_{i\perp}$ \citep{Phan2013ElectronShear,Haggerty2015TheReconnection}. This can be rearranged as $\Delta T_{i\parallel} - \Delta T_{i\perp} \approx 0.094 m_i c_{Ar}^2$. Substituting in the difference in heating we find that firehose instability should be approximately suppressed for $B_g \approx  0.43 B_r$ which corresponds to an approximate shear angle of $135^{\circ}$. This value is consistent with the transition found in simulations and serves as a natural value to separate nearly anti-parallel and guide field reconnection.

\section{\label{sec:Disc} Discussion and Conclusion}
We have shown in simulations and observations that there is a systematic reduction of the outflow velocity in nearly anti-parallel magnetic reconnection events. The reduction of the outflow velocity is correlated with the ion temperature and is shown to be due to the firehose instability in the exhaust. The outflow velocity is shown to be well predicted by $v_0 = c_{Ar}^2/2\sqrt{T_{i\parallel}/m_i}$. It is also shown that for events with a sufficiently strong guide field (with strength comparable to the reconnecting magnetic field) that the outflow velocity reaches the Alfv\'en speed.  The clear agreement between the theory proposed in this paper with the simulations and observations strengthens the claim that the firehose instability in reconnection exhausts is responsible for the reduction of the outflow velocity. 

This result has significant implications for an important open question about the nature of collisionless magnetic reconnection: What is the partition of converted magnetic energy? The bulk outflow contains a significant fraction of the released magnetic energy \citep{Birn2010ScalingPlasmas,Shay2014ElectronStudy,Eastwood2013EnergyMagnetotail,Yamada2014ConversionPlasma} and so if the outflow velocity is reduced and the total magnetic energy released remains the same then more energy will be released into other degrees of freedom. Therefore, the relationship between exhaust ion temperature and outflow velocity described here is an important step towards a predictive model of the partition of reconnection energy.


\begin{acknowledgments}
This research was supported by NSF grant AGS-1460130 and AGS-1460037 as well as NASA grants
NNX08A083G (MMS IDS), NNX14AC39G (MMS Theory and Modeling team), NNX15AW58G, NNX16AF75G, NNX16AG76G, and NNX17AI25G and Thailand Research Fund grant RTA5980003.
Simulations and analysis were performed at the National Center for Atmospheric Research Computational
and information Systems Laboratory (NCAR-CISL) and at the National Energy Research Scientific
Computing Center (NERSC). Rungployphan Kieokaew acknowledges financial support from Faculty of Science at Mahidol University on short-term research overseas to visit Bartol Research Institute.
\end{acknowledgments}

\newpage
\bibliography{Mendeley,bib}
\end{document}